\documentclass[letterpaper]{article} 
\usepackage{aaai2026}  
\usepackage{times}  
\usepackage{helvet}  
\usepackage{courier}  
\usepackage[hyphens]{url}  
\usepackage{graphicx} 
\usepackage{natbib}  
\usepackage{caption} 
\usepackage{algorithm}
\usepackage{algorithmic}

\usepackage{booktabs}
\usepackage{tabularx}
\usepackage{amsmath}
\usepackage{amssymb}
\usepackage{siunitx}
\usepackage{ragged2e}

\usepackage{soul}
\usepackage{enumitem}
\usepackage{multirow}
\usepackage{longtable}

\usepackage{xcolor}

\definecolor{autogluonblue}{HTML}{005AB5}
\definecolor{xgbred}{HTML}{DC3220}

\usepackage{eso-pic}
\usepackage{xcolor}
\newcommand{\copyrightbanner}{
    \AddToShipoutPictureFG*{
        \AtPageUpperLeft{
            \raisebox{-1.5cm}{
                \makebox[\paperwidth][c]{
                    \color{red}
                    \parbox{0.6\paperwidth}{
                        \centering\bfseries\small
                        Please check and cite the published version of this paper in the Proceedings of the 20th International AAAI Conference on Web and Social Media (ICWSM 2026)
                    }
                }
            }
        }
    }
}

\usepackage{newfloat}
\usepackage{listings}
\DeclareCaptionStyle{ruled}{labelfont=normalfont,labelsep=colon,strut=off} 
\floatstyle{ruled}
\newfloat{listing}{tb}{lst}{}
\floatname{listing}{Listing}
\title{Self-Moderation in the Decentralized Era: Decoding Blocking Behavior on Bluesky}
\author {
    Carlo Alberto Bono\textsuperscript{\rm 1},
    Nick Liu\textsuperscript{\rm 2},
    Giuseppe Russo\textsuperscript{\rm 3},
    Francesco Pierri\textsuperscript{\rm 1}
}
\affiliations{
    \textsuperscript{\rm 1}Politecnico di Milano, Milan, Italy \\
    \textsuperscript{\rm 2}Indiana University, Bloomington, IN, USA \\
    \textsuperscript{\rm 3}EPFL, Lausanne, Switzerland \\
    carlo.bono@polimi.it
}

\begin{document}

\maketitle
\copyrightbanner

\begin{abstract}
Centralized social media platforms combine top-down moderation with user-level mechanisms like blocking, but access to behavioral and moderation-related data is typically restricted. 
In contrast, emerging decentralized networks like Bluesky provide researchers with open access to blocking actions, offering a unique opportunity to study self-moderation at scale.
Understanding blocking behavior is critical in order to assess how community safety and user autonomy can be balanced in these environments. 
This study examines blocking on Bluesky, analyzing more than 100M actions by nearly 2M users over three months. 
We construct behavioral profiles from 86 features capturing user activity, content, and network interactions, and address two questions: (1) Can a user’s propensity to be blocked by others be estimated from in-platform behavior? and (2) Which behavioral features are most informative in predicting this propensity? 
Using state-of-the-art machine learning models, we achieve high accuracy in binary classification (AUROC $>$ 0.85) and reliable performance in regression ($R^2 \approx 0.5$). 
Our findings show that a small subset of features is sufficient for robust prediction, even when accounting for higher activity on the platform. 
Through explainability analyses, we identify the behavioral signals most strongly associated with blocking outcomes, enabling a transparent understanding of why certain users are more likely to be blocked. 
Our study provides a framework and empirical evidence that advances understanding of both the mechanisms and feature signals underlying user self-moderation, while also offering insights relevant to moderation design. 
\end{abstract}

%

\section{Introduction}
The rise of decentralized social platforms has caused significant shifts in the way online communities are managed, evolve and interact~\cite{Zignani2018FollowT}. Unlike traditional social networks, decentralized platforms prioritize the principles of user autonomy and control, allowing for a more personalized and attuned experience~\cite{Raman2019ChallengesIT}. Emerging decentralized platforms have gained attention for their innovative approach to social networking and their potential to reshape online interactions, with mechanisms similar to blockchain technology~\cite{guidi2020blockchain}. Decentralized approaches enable independent servers to host content, reducing the risk of centralized control and empowering users with greater autonomy over their data and content~\cite{Raman2019ChallengesIT,Zignani2018FollowT}. 
However, together with the freedom offered by a decentralized paradigm, also comes the challenge of moderating harmful and abusive behavior~\cite{ghenai2025exploring,zhang2024predicting,bonifazi2022investigating,bono2024exploration}. Online abuse, harassment, and toxic interactions have already been longstanding issues on centralized platforms, prompting various moderation strategies over time, ranging from algorithmic content filtering to user reporting systems~\cite{jhaver2021evaluating}. In centralized social networks, moderation is usually enforced through top-down and platform-driven mechanisms~\cite{chandrasekharan2017you,pierri2024drivers,di2022vaccineu}. Decentralized platforms, where content moderation is often less centralized, rely more on user-driven moderation, with blocking actions serving as a tool for managing unwanted interactions~\cite{Zignani2018FollowT}.

On online social platforms, users may become aware that they have been blocked, possibly through indirect indicators.
On microblogging platforms, blocking a user prevents any form of interaction and removes their presence from the experience of the blocking user.
Blocked users are unable to like, reply to, mention, or follow the blocker, and their posts, replies, and profile are no longer visible to the blocking user.
However, information about block actions is not always publicly accessible.
The approach of Bluesky, one of the emerging decentralized social platforms~\cite{balduf2024bluesky,balduf2025bootstrapping,failla2024m,sahneh2024dawn}, is distinctive in that user blocks are currently public and enumerable, since all servers across the network must be aware of blocks in order to honor them. This approach presents a novel opportunity for research into user behavior and moderation dynamics within decentralized social networks. This transparency introduces both positive and negative potential implications. On one hand, users could be aware of possibly problematic interactions, but on the other hand, social dynamics could be influenced as blocks may carry stigmas or lead to public shaming. As a result, understanding the behavioral features that predict the propensity to be blocked is critical from the perspective of both users and platform developers, and researchers.


While a few studies have analyzed behaviors such as muting and unfollowing on social media through experiments~\cite{martel2024blocking,rathje2024unfollowing}, these actions have generally been difficult to study in detail, since platforms do not make this kind of user activity accessible. This paper aims to fill this gap by examining the behavioral features that correlate with being blocked, and investigating the extent to which behavioral patterns can predict being blocked. Our analysis seeks to contribute to the broader discussion on online moderation and user behavior in decentralized social networks. 
In doing so, we place particular emphasis on model interpretability, which enables us to identify and characterize behavioral patterns most strongly associated with blocking.

We investigate the following research questions:

\begin{itemize}[labelwidth=2em, labelsep=0.5em, leftmargin=!] 
\item[\textbf{RQ1}] Can we estimate Bluesky users' propensity to be blocked based on their observed in-platform activity?
\item[\textbf{RQ2}] Which behavioral features are most indicative of a user’s propensity to be blocked?
\end{itemize}

To this end, we collect a longitudinal dataset of all the Bluesky activity over a three-month period (June-August 2024) and extract over 80 features that describe online behaviour from different perspectives, such as activity, content, and interactions. 
We frame the task as both a binary classification and a regression task to assess the predictability and intensity of blocking behavior within the observation period, in a retrospective manner.
We also apply feature importance techniques to quantify the informativeness of features and feature groups for detecting blocked users. 
Our findings show that extremely blocked users can be identified with high accuracy, offering insights that can inform the design of moderation tools and interventions.

\section{Related Work}
\label{sec:related_work}
\subsection{Centralized and Decentralized Moderation Policies}
Research on platform moderation has investigated the behavioral and ecological effects of interventions, typically distinguishing between ``hard'' and ``soft'' moderation strategies \cite{trujillo2022make, schneider2023effectiveness}. Hard moderation, such as banning communities or users, directly removes content or accounts \cite{young2022much, rogers2020deplatforming}, while soft moderation, including visibility adjustments or contextual interventions, reshapes user behavior without removal \cite{shen2022tale, zannettou2021won,attanasio2026effects}. Reddit’s quarantining mechanism is an example of soft moderation, limiting community growth while retaining activity, though it has been criticized for reinforcing echo chambers and allowing platforms to profit from controversial content \cite{copland2020reddit, chandrasekharan2022quarantined}. Community-driven initiatives like X’s Community Notes similarly aim to contextualize potentially misleading content, increasing trust in fact-checking but struggling to prevent initial exposure to misinformation \cite{chuai2023roll, drolsbach2024community, solovev2025references}.
Banning, a common hard moderation approach, removes harmful communities entirely \cite{innes2023platforming, ali2021understanding}. While effective at reducing harmful content on mainstream platforms, deplatformed users often migrate to fringe platforms, potentially reinforcing toxic communities and facilitating spillover of harmful behavior \cite{horta2021platform, zuckerman2021deplatforming}. Targeted moderation at the individual level—deplatforming influencers \cite{ribeiro2024deplatforming} or mitigating ``superspreaders'' of misinformation \cite{baribi2024supersharers}—can limit harm effectively without the broader disruptions caused by community-wide interventions.
Decentralized platforms have drawn attention for their transparency and user-driven moderation mechanisms \cite{zia2023flocking, Raman2019ChallengesIT}. For example, Bluesky has open-sourced Ozone, its official moderation tool, enabling users and developers to create independent moderation services and customizable content filters, expanding the ways in which moderation and self-curation can be implemented beyond individual blocklists~\cite{balduf2024bluesky, kleppmann2024bluesky}, while prior work on Mastodon highlights how decentralized moderation often relies on on server-level blocklisting practices~\cite{bono2024exploration}.
This multi-tiered approach highlights the potential of community-involved moderation in decentralized contexts.

\subsection{Bluesky in Research}
Recent studies of Bluesky have characterized its architecture, user activity, and unique features. \citet{balduf2024bluesky} provide a large-scale analysis of Bluesky’s modular architecture and third-party provider ecosystem, while \citet{failla2024m} release a high-coverage dataset encompassing 4 million users and 235 million posts, including social interactions and content recommendation outputs. \citet{quelle2025bluesky} examine user-generated algorithmic feeds, noting widespread creation but limited adoption, with polarization emerging on topics such as the Israel-Palestine conflict. \citet{sahneh2024dawn,nogara2026longitudinal} longitudinally analyze activity during Bluesky’s public rollout, observing patterns similar to established platforms, higher volumes of original content compared to resharing, and low toxicity levels, suggesting effective moderation during rapid growth. 
These studies collectively illustrate Bluesky’s value as a platform for examining decentralized social dynamics.

\subsection{Self-Moderation in Online Social Platforms}
Self-moderation mechanisms, such as blocking or muting, allow users to reduce exposure to unwanted content or harassment \cite{vogels2021state,seering2020reconsidering}. Research indicates that user blocking is often politically or ideologically motivated, with users disproportionately blocking counter-partisan accounts or misinformation spreaders, thereby increasing network polarization \cite{kaiser2022partisan, martel2024blocking, baysha2020dividing}. Studies also highlight the persistence of exposure to misinformation despite blocking or unfollowing behaviors, pointing to the limits of reactive strategies \cite{ashkinaze2024dynamics}.
Psychological and social factors further influence blocking behaviors. Blocking can protect mental well-being \cite{hunt2018no, fox2015dark}, reflect cultural norms \cite{zhang2020privacy}, and be guided by peer influence. 

\subsection{Position of Our Work}
Prior work has examined self-moderation through surveys, experiments, or small-scale analyses, but has not tested whether users’ behavioral traces can systematically explain or predict blocking outcomes. 
Bluesky’s public access to self-moderation data provides a rare opportunity to analyze these behaviors at scale, offering insights into dynamics that were previously difficult to study on centralized platforms.
Our study addresses this gap by formulating the problem of blocking behaviour as both a binary classification and a regression task, applying state-of-the-art models to Bluesky’s publicly accessible self-moderation data. 
By evaluating the predictive power of behavioral features in a retrospective setting, our work provides the first large-scale evidence that blocking behavior can be inferred from per-user activity patterns, offering a foundation for future efforts to anticipate blocking events and to investigate their causal effects on user behavior.

\section{Dataset and Methods}
\label{sec:dataset}

\begin{figure}[!t]
    \centering
\includegraphics[width=0.7\linewidth]{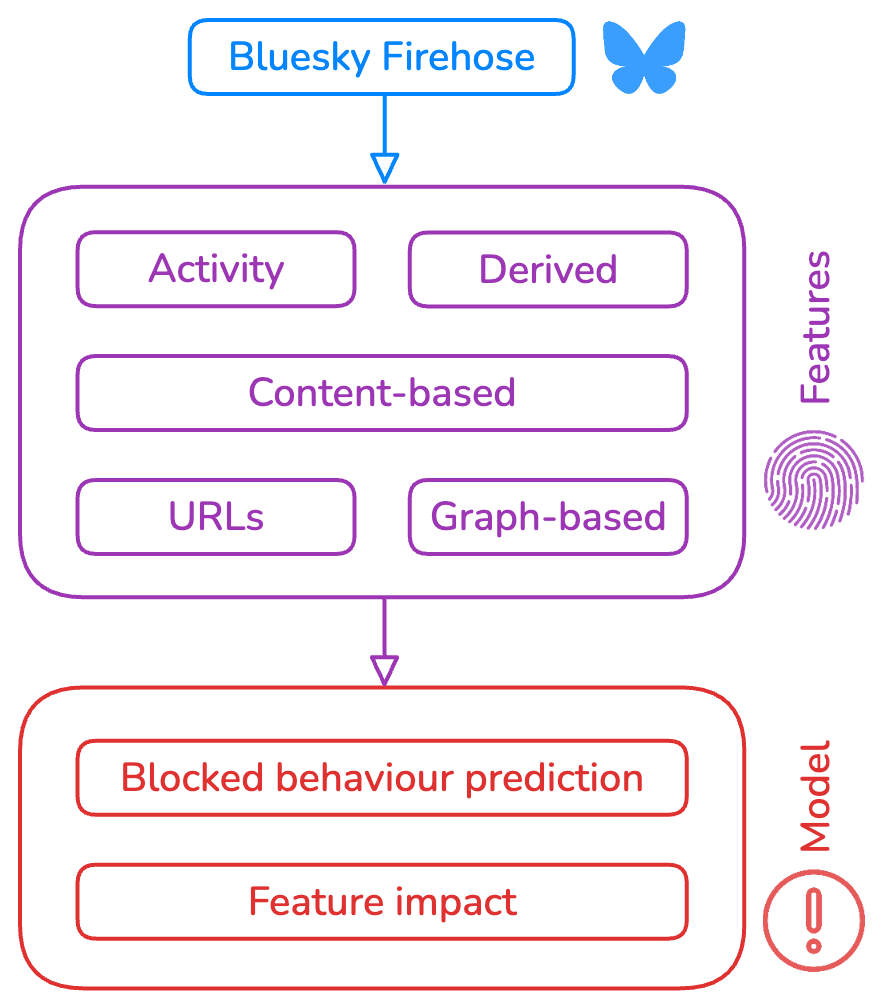}
    \caption{Diagram describing the workflow of our research.}
    \label{fig:diagram}
\end{figure}

The methodology employed in this study is outlined in Figure~\ref{fig:diagram} and can be summarized as follows.
We first collect all Bluesky public data within a specified timeframe. 
We then build user profiles for all the active users observed, incorporating several characteristics of their activity, interactions, and shared content. 
To evaluate whether these behavioral signals explain blocking outcomes, we formulate the task as both a binary classification problem and a regression problem, and we leverage state-of-the-art machine learning methods to assess the informational value of these features in relation to users’ propensity to be blocked by others.

\subsection{Data Collection}

We collected data continuously over a period of roughly three months, from June 1st, 2024, to August 28th, 2024. Statistics for the resulting dataset are provided in Table~\ref{tab:dataset-summary}.

Data has been obtained through Bluesky’s publicly accessible Firehose endpoint, which provides unauthenticated, real-time access to granular platform activity. A key feature of the endpoint is the ability to resume data retrieval after connection interruptions, allowing data collectors to recover missing records from up to 72 hours prior. This ensures a continuous data stream throughout the observation period, mitigating potential gaps caused by temporary network failures. To perform data acquisition, we utilized the \texttt{bluesky} Dart library \citep{BlueskyDart2024} and an established open-source library \citep{BurghardtFirehose2024} to interact with the AT Protocol ecosystem and consume the \texttt{com.atproto.sync.subscribeRepos} endpoint --- commonly known as the Firehose\footnote{\url{https://docs.bsky.app/docs/advanced-guides/firehose}} stream \citep{BlueskyFirehose2024}. 

Bluesky enables the monitoring of public user activities, which offers valuable insights into social interactions within the platform. Firehose data include user-generated content such as posts, replies, follows, likes, and other interactions ~\citep{bluesky_firehose, atProtocol_eventStream}. Our study considers five primary forms of interaction on the platform:

\begin{itemize}
    \item \textbf{Posts} – Original content shared by users.  
    \item \textbf{Replies} – Direct responses to existing posts, fostering discussion.  
    \item \textbf{Reposts} – Sharing of content originally published by other users.  
    \item \textbf{Follows} – Directed social connections where one user subscribes to another user’s updates. 
    \item \textbf{Blocks} – Symmetric user actions restricting content visibility and interactions among specific accounts.  
\end{itemize} 

Specifically, blocking a user restricts all forms of interaction and removes their presence from the experience of the user performing the block. 
A blocked account cannot like, reply to, mention, or follow the blocking user.
Additionally, their posts, replies, and profile will be hidden from search results. 
Unlike other platforms, blocks on Bluesky are public\footnote{\url{https://docs.bsky.app/blog/block-implementation}}, enabling the study of the propensity of being blocked by other users. 
Mute operations, which are asymmetrical operations causing users' posts to be excluded from a feed, are instead private on Bluesky and are not included in this study.

Although Bluesky’s terms of service do not impose restrictions on data collection from publicly accessible content, we adhere to strict ethical guidelines. Our dataset comprises only publicly available data collected in accordance with the platform’s Privacy Policy.\footnote{\url{https://bsky.social/about/support/privacy-policy}} To uphold user privacy, we do not make any raw data available. Instead, this paper presents only anonymized, aggregate observations.

\begin{table}[!t]
\centering
\caption{Summary of the collected dataset (June 1 -- August 28, 2024).}
\begin{tabular}{@{}lr@{}}
\toprule
Item & Count \\
\midrule
Blocks         & \num{3278406} \\
Follow actions & \num{33942018} \\
Likes          & \num{292388501} \\
Posts          & \num{79737148} \\
Reposts        & \num{30284394} \\
Unique users   & \num{1979713} \\
\bottomrule
\end{tabular}
\label{tab:dataset-summary}
\end{table}


\subsection{Feature definition and preprocessing}
\label{sec:features}
In our analysis, we focus on the subset of users who published at least 10 posts during the observation period, irrespective of the account creation date and whose posts are most frequently written in English, resulting in a total of \num{427118} users. For each of these users, we extract 86 numerical features covering multiple aspects of their behavior. 
These features are derived from both the collected data and external sources, providing a comprehensive view of user activity during the observation period.
The complete list of features, with descriptions and logical groupings, is provided in the Appendix.
Taken together, these features aim to capture meaningful behavioral dimensions that distinguish different types of users, such as high-frequency posters, socially central accounts, or users who share more toxic or politically polarized content.
This breadth enables us to characterize user profiles in a way that is interpretable and behaviorally grounded, helping to contextualize why some users may be more likely to be blocked.
Features related to network centrality and hyper-activity have, for instance, been linked to bot-like or abusive users on Bluesky~\cite{nogara2026longitudinal}.

\noindent{\textbf{Action Features}} These 18 features are obtained by counting the number of relevant events observed in the dataset. They include counts for the following actions types: likes, posts, reposts, follows, and blocks. Each event can be a create or a delete action, which are counted separately. Moreover, since the source and target user handles are known, they are used to calculate \textbf{Derived} counts for actions directed at each user, that is, the number of events in which a user has been liked, reposted, followed, or blocked. The counts of replies, as an author or as a subject, are derived from the post contents.


\noindent{\textbf{Post-derived Features}} The following 13 textual features are extracted from the content of each post: total number of characters, number of lowercase and uppercase characters, digits, spaces, and emojis. We compute the mean and standard deviation of these features aggregating them at the user level. We also compute the variability of the declared post languages as the normalized Shannon entropy of the observed languages. 

We associate several toxicity dimensions (Identity attack, Insult, Obscene, Toxicity, Severe toxicity, Threat, Sexually explicit) with each post using Detoxify~\cite{Detoxify}, restricting the application to posts written in its supported languages\footnote{`en', `pt', `ru', `es', `fr', `tr', `it'.}. Multilingual toxicity classification models can, in general, show variable performance across languages. This observation is partially mitigated by the fact that, in our experimental setup, user-level toxicity aggregates are dominated by English content, for which Detoxify has been extensively validated\footnote{We recall that users included in the analysis are required to have English as the most frequent language of their posts.}.

The mean and standard deviation of these toxicity indicators are calculated at the user level, resulting in 14 features.
Each dimension has a value in the interval $(0, 1)$.
Since our analysis only includes users with at least 10 posts, no missing values are present in the data.

Additionally, when posting original content, users may include URLs in their posts. We extract all the URLs contained in the posts and parse the domain names from these URLs. For each user, we calculate the average number of URLs per post and the overall variability of the domains, again as the normalized Shannon entropy of the observed domains.

The computation of all the post-derived features considers only original post objects, excluding reposts.

\noindent{\textbf{URL-derived Features}} We utilize the domains of the shared URLs to characterize users in terms of misinformation and credibility For each user, we count the number of domains shared, grouped according to the categories defined by Media Bias/Fact Check (MBFC\footnote{\url{https://mediabiasfactcheck.com/}}), considering dimensions of \textit{bias} (Extreme-left, Left, Center-left, Center-right, Right, Extreme-right, Satire, Conspiracy, Pro-science), \textit{credibility} (Low, Medium, High), and \textit{factuality} (Very low, Low, Medium, Mostly, High, Very high, Mixed), resulting in 19 features. These counts are then normalized by the total number of posts published by the user. Additionally, we compute the average domain quality score based on the ratings provided by \cite{Lin2023Sep}, substituting missing values with the average of the valid scores. The number of shared domains matching these two external sources is also used as a feature, again normalized by the number of posts.

\noindent{\textbf{Graph-based Features}} Centrality measures are widely used to capture users' influence and position within a network~\cite{Newman2010}. We compute graph-based centrality measures --- Coreness, total Degree, and PageRank --- from the networks of the following interactions: follows, likes, replies, and reposts, considered separately. Coreness measures the depth of a node within the network, Degree quantifies the number of direct connections a user has, and PageRank evaluates a user's importance based on the probability that a random walk over the network lands on them.

\subsection{Target variables and predictive tasks}
\label{sec:target}
We formalize blocking prediction as both a binary classification task and a regression task, enabling us to examine whether users’ behavioral features can explain not only the occurrence but also the intensity of blocking behavior.
We remark that this analysis is conducted retrospectively on existing blocking events, and not to predict future blocks, which we leave for future work.

Since no standard threshold exists for classifying a user as \textit{blocked}, we consider two target variables to capture blocking behavior: the absolute number of blocks ($raw_{blocked}$) and the activity-normalized number of blocks ($norm_{blocked} = raw_{blocked} \,\scalebox{0.9}{$/$}\, \scalebox{0.9}{$\#$}posts$). 
The normalization is motivated by the fact that highly active users may appear to be blocked more often simply due to their overall activity level. 
We remark that we focus only on active users who shared at least 10 posts over the observation period (\num{427118}). 

\noindent\textbf{Task 1: Binary Classification} The first task is a binary classification problem: using percentiles of the two variables defined above ($raw_{blocked}$ and $norm_{blocked}$) as thresholds, we predict whether a user belongs to the positive \textit{blocked} class.  

We employ Random Forest classifiers as tree-based ensemble methods are well-suited for heterogeneous, non-linear feature spaces. We implement them via \texttt{XGBoost} with 500 estimators, evaluated with the average of 10 independent runs of 10-fold cross-validation, with the label distribution balanced by randomly undersampling the majority class.  This procedure aims at a greater robustness, particularly at higher thresholds where the label imbalance becomes more extreme.
Feature importance is quantified with SHAP \citep{Lundberg2017Unified}, which provides model-agnostic, locally consistent attributions and allows us to interpret how individual behavioral features contribute to predictions. Both XGBoost and SHAP value analyses are state-of-the-art approaches for modeling and interpreting user behavior in social media settings~\cite{shevtsov2022identification,mathew2019thou,yang2020scalable}.
Performance is measured with Receiver Operating Characteristic Area Under the Curve (ROC AUC), where 0.5 indicates random predictions and 1 perfect classification.
    
\noindent\textbf{Task 2: Regression} The second task is a regression problem, where we obtain a model estimate $\hat{y}$ of the actual number of blocks received $y$, according to both raw and activity-normalized definitions. 
We use Random Forest Regression (XGBRFRegressor) and AutoGluon TabularPredictor \citep{agtabular} with the ``high quality'' setting. 
AutoGluon’s automated model selection and ensembling provide strong performance on heterogeneous tabular data with minimal tuning. 
Model performance is assessed using $R^2$ and Mean Absolute Error (MAE):
\[
R^2 = 1 - \frac{\sum_{i=1}^{n} (y_i - \hat{y}_i)^2}{\sum_{i=1}^{n} (y_i - \bar{y})^2}, \quad 
\text{MAE} = \frac{1}{n} \sum_{i=1}^{n} |y_i - \hat{y}_i|,
\]
using an 80:20 random train-test split.

\section{Results}
\label{sec:results}

\subsection{Exploratory analysis}

\begin{figure}[!t]
    \centering
\includegraphics[width=\linewidth]{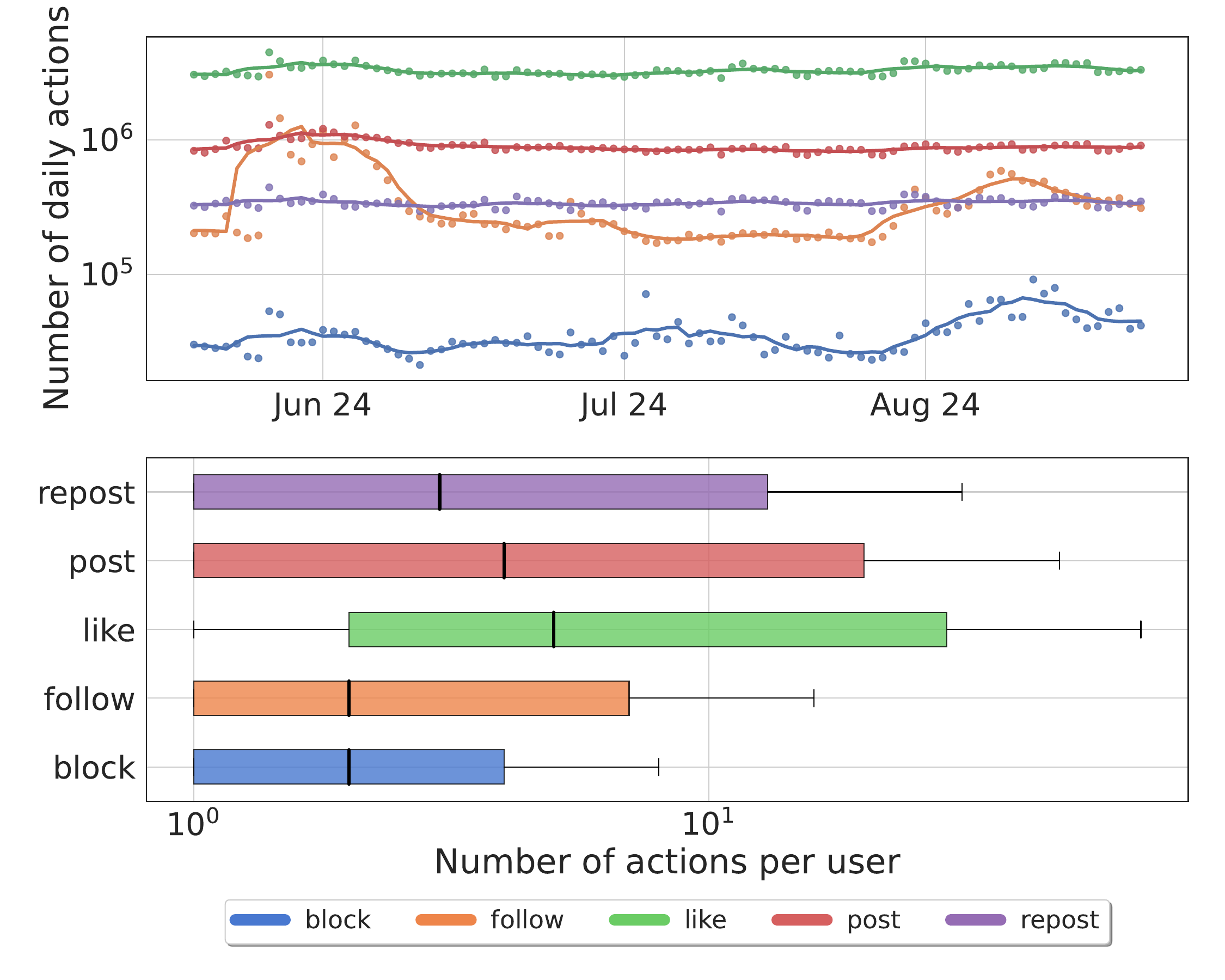}
    \caption{\textbf{(top)} Time series of the number of daily actions performed on Bluesky during the observation period. Daily observations and a 7-day moving average are reported. The scale of the y-axis is logarithmic. \textbf{(bottom)} Boxplot of the number of actions (log scale) performed by Bluesky users. Median values of each distribution in the bottom plot are: blocks = 2, follows = 2, likes = 8, posts = 6, reposts = 4.}
\label{fig:descr_timeseries}
\end{figure}

\begin{figure}[!t]
    \centering
\includegraphics[width=\linewidth]{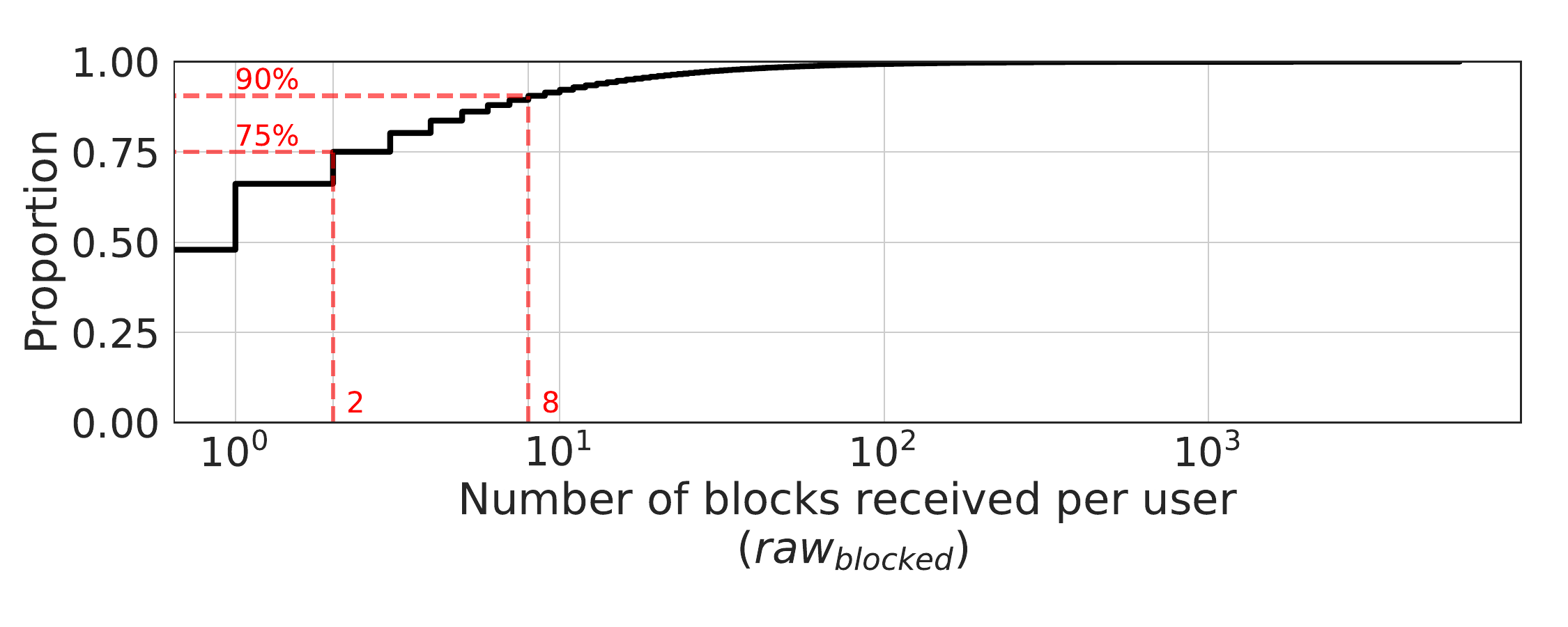}
    \caption{Empirical cumulative distribution of the number of blocks received by a user (log scale), considering users that shared at least 10 posts over the observation period.}
    \label{fig:blocked}
\end{figure}

\begin{figure}[!t]
    \centering
\includegraphics[width=\linewidth]{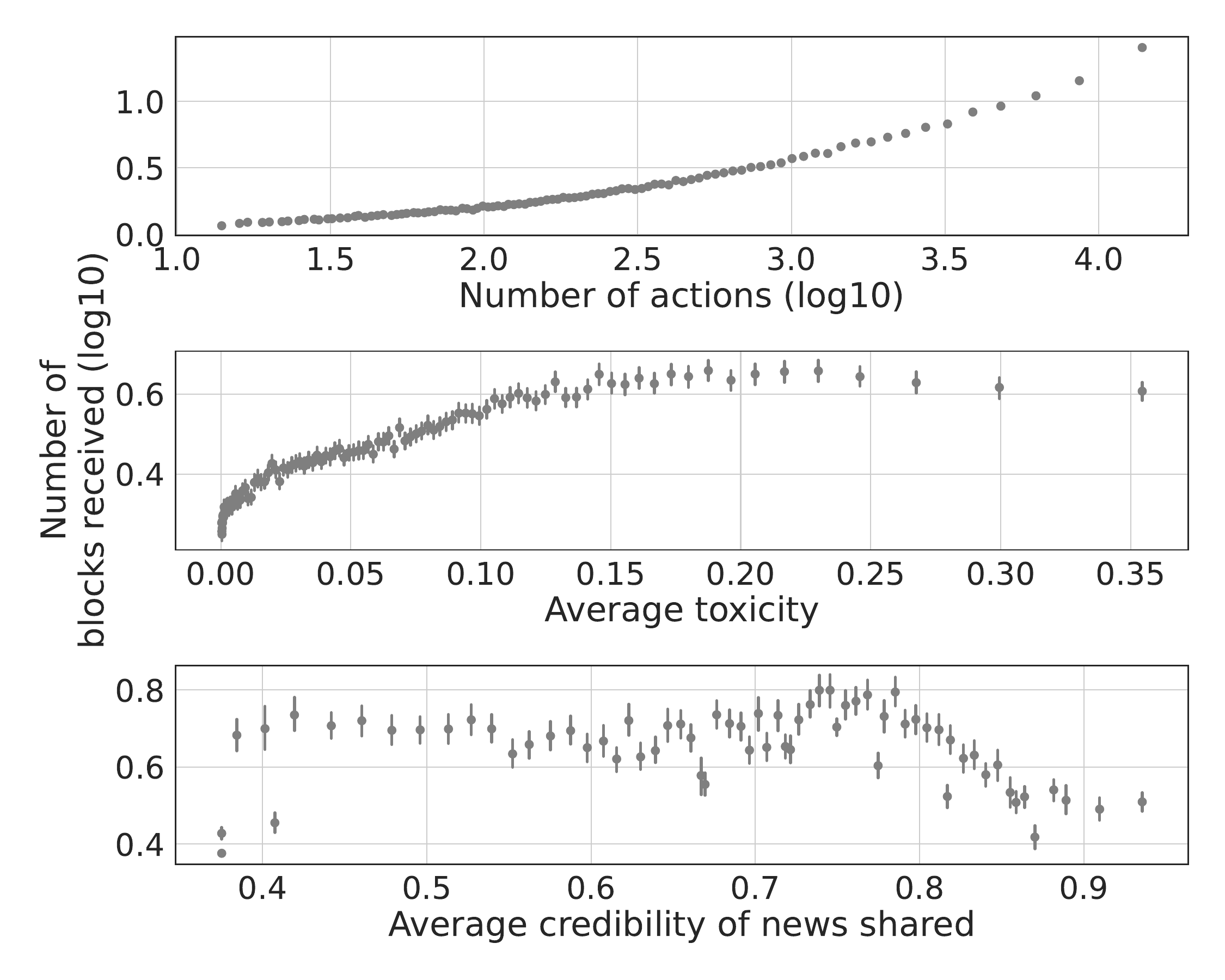}
    \caption{Correlation between different user features --- number of actions, average toxicity and average credibility of news shared --- and the number of blocks received ($\text{log}_{10}$). We bin the \textit{x} variable into 100 discrete bins and then estimate the mean value of the \textit{y} variable with 95\% confidence interval using bootstrapping. The analysis is performed on \num{427118} users sharing at least 10 posts.}
    \label{fig:descr_correlations}
\end{figure}
We first provide descriptive statistics of the dataset, offering an overview of Bluesky users' activity during the period of analysis.

The top panel of Figure~\ref{fig:descr_timeseries} reports the temporal distribution of actions performed by Bluesky users during the observation period. Likes are the most frequent action, with a daily median of \num{3240712}. Original posts occur nearly three times as often (daily median = \num{873760}) than reposts (daily median = \num{338544}). Users also frequently follow each other, with a daily median of \num{248255} follow actions. In contrast, block actions are significantly less common, with a daily median of \num{31934}, making them several orders of magnitude less frequent than the other interactions. In terms of temporal trends, we observe a spike in follow actions in June, likely driven by a surge in the platform’s popularity. Additionally, an increasing trend in follow and block actions can be observed toward the end of August, which can be attributed to growing political engagement ahead of the 2024 U.S. Presidential elections.

Bottom panel of Figure~\ref{fig:descr_timeseries} illustrates the distribution of user actions on Bluesky during the observation period. As expected in online social networks, all distributions exhibit a power-law behavior, where the majority of users perform only a limited number of actions, while a small fraction of users exhibit significantly higher activity levels, with extreme outliers carrying out up to \num{74000} blocks, \num{517000} likes, \num{145000} follow actions, \num{316000} original posts, and \num{82000} reposts. 
This suggests the presence of potentially misbehaving users, likely engaging in harassment, spam, or automated activity on the platform. 

Our primary objective is to investigate and predict the propensity of being \textit{blocked} on Bluesky. As specified in the Methods, we focus on over 400k users who shared at least 10 posts over the observation period. Figure~\ref{fig:blocked} shows the distribution of the number of blocks received by these users. We can observe that the large majority of users ($90\%$) received less than 10 blocks, with a small but non-negligible proportion of users that received over 100 blocks ($\approx0.6\%$). Figure~\ref{fig:descr_correlations} illustrates the relationship between the number of blocks ($\text{log}_{10}$ scale) received by users and three features for user characterization: total actions performed, average toxicity of posted messages and average credibility of shared news URLs. A strong positive association, likely exponential given the logarithmic scale, between a user’s total activity on the platform and the number of blocks received can be observed, suggesting that highly active users are more likely to be blocked (Pearson's $R=0.55$, Spearman's $R=0.53$), potentially due to misbehavior, spamming, or as a result of their increased visibility.
Similarly, the average message toxicity of users is moderately correlated with the number of blocks received (Pearson's $R=0.20$, Spearman's $R=0.23$). 
In contrast, no clear relationship emerges between sharing less credible news sources and being blocked.

\begin{figure}[!t]
    \centering
    \includegraphics[width=\linewidth]{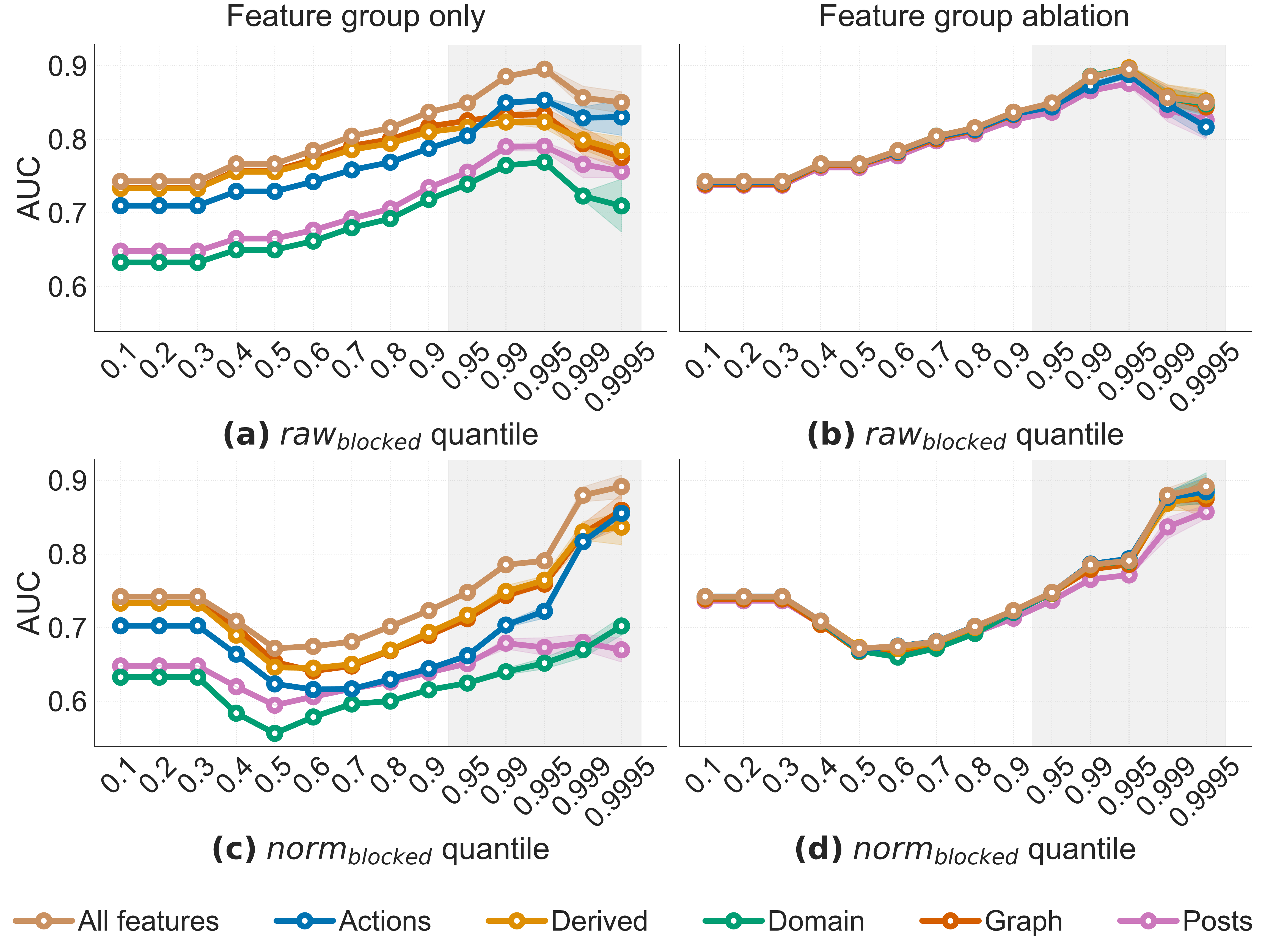} 
    \caption{ROC AUC of the XGBoost classifier on the positive class (\textit{blocked} users) for different quantiles of the \textbf{(a)} raw and \textbf{(c)} normalized block count, trained using \textit{only} one subset of features at a time. Panels \textbf{(b, d)} show the performance when training models using all \textit{except} one subset of features at a time. All panels also report the ROC AUC of a predictor trained with all the features, for comparison purposes. Labels in the legend indicate groups of features as defined in the Methods. The grey area highlights performance for the thresholds in the range $[0.9, 0.9995]$.} 
    \label{fig:classify}
\end{figure}

\subsection{Classification task}
\label{sec:classification}
As specified in the Methods, we formulate a binary classification task to assess how well \textit{blocked} and \textit{non-blocked} users can be distinguished. We perform the same experiments independently for the $raw_{blocked}$ and $norm_{blocked}$ measures. Users below a threshold are labeled as the negative class (\textit{non-blocked}), while those above are labeled as the positive class (\textit{blocked}). To evaluate the informativeness of features for this task, we train classifiers on different subsets of features, grouped by type. 

Panels~\textbf{a, c} of Figure~\ref{fig:classify} show the performance, in terms of ROC AUC, of the binary classifiers across different thresholds for $raw_{blocked}$ and $norm_{blocked}$, respectively. 
We evaluate a classifier for each feature group and compare its performance against using all features. 
Results are averaged over 10 different runs for each classifier, each run consisting of an independent random 10-fold cross-validation.
We first observe that the classifier trained on all features consistently outperforms reduced models across thresholds in both settings, reaching max AUC values of \num{0.892} and \num{0.875}, respectively. Performance improves with larger thresholds, though with distinct patterns. In the raw block count setting, AUC steadily increases up to the highest thresholds, after which it slightly declines (quantiles $0.995–0.9995$). In contrast, in the normalized block count setting, performance dips at mid-range quantiles ($0.3–0.7$) before rising sharply, reaching its peak at the highest quantile ($0.9995$). These trends suggest that more extreme users are generally easier to distinguish, particularly when block counts are normalized by activity.
In the same panels, we also observe that groups of behavioral features like  Action, Derived, and Graph, are highly effective at predicting blocked users even when used in isolation, achieving AUC values up to \num{0.856} and \num{0.846} in the $raw_{blocked}$ and $norm_{blocked}$ settings respectively, and mimicking the overall trend of the classifier trained on all features. In contrast, features like Domain and Posts appear to be less informative, with the weakest performance observed in the $norm_{blocked}$ setting, where their AUC remains below 0.7.

\begin{figure}[!t]
    \centering    \includegraphics[width=\linewidth]{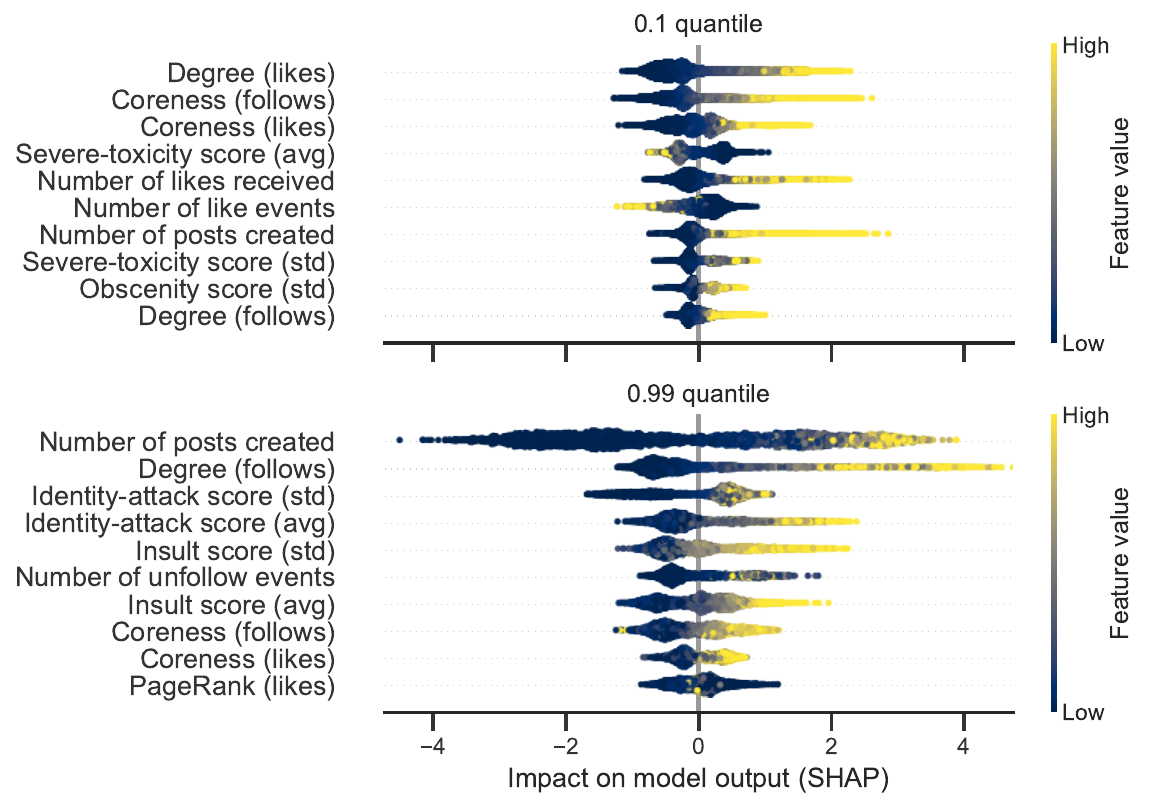} 
    \caption{SHAP beeswarm plots comparing the 0.1 and 0.99 quantile setups, top 10 features reported.}
    \label{fig:shap_beeswarm_quantiles}
\end{figure}

To gain insight into the contribution of individual features to blocking, Figure~\ref{fig:shap_beeswarm_quantiles} reports SHAP beeswarm plots for two representative thresholds (quantiles 0.1 and 0.99), showing the top 10 most influential features for each model. It is worth noting that, when focusing on more extreme users with the higher quantile, substantially larger SHAP magnitudes indicate stronger and more polarized feature effects, in line with the previous observations.

We further conducted a feature ablation study, training binary classifiers over the same thresholds while removing a single feature group at a time. As shown in panels~\textbf{b, d} of Figure~\ref{fig:classify}, most settings yield performance comparable to the full model, with the absence of Action or Post features causing the largest decline ($\approx-5$ p.p.). These results highlight a non-trivial interplay among features: for example, post-related features alone have limited predictive power, yet their removal from the full set produces a noticeable drop in performance. This effect is most evident at higher quantiles, while at lower quantiles, the absence of a feature group appears to be compensated by the remaining features. We examine the contribution of individual features in more detail in the next section.

\subsection{Feature importance analysis}
We investigated the importance of individual features in driving the classification performance using SHAP analysis as described in the methods. 

\begin{figure}[!t]
    \centering    \includegraphics[width=\linewidth]{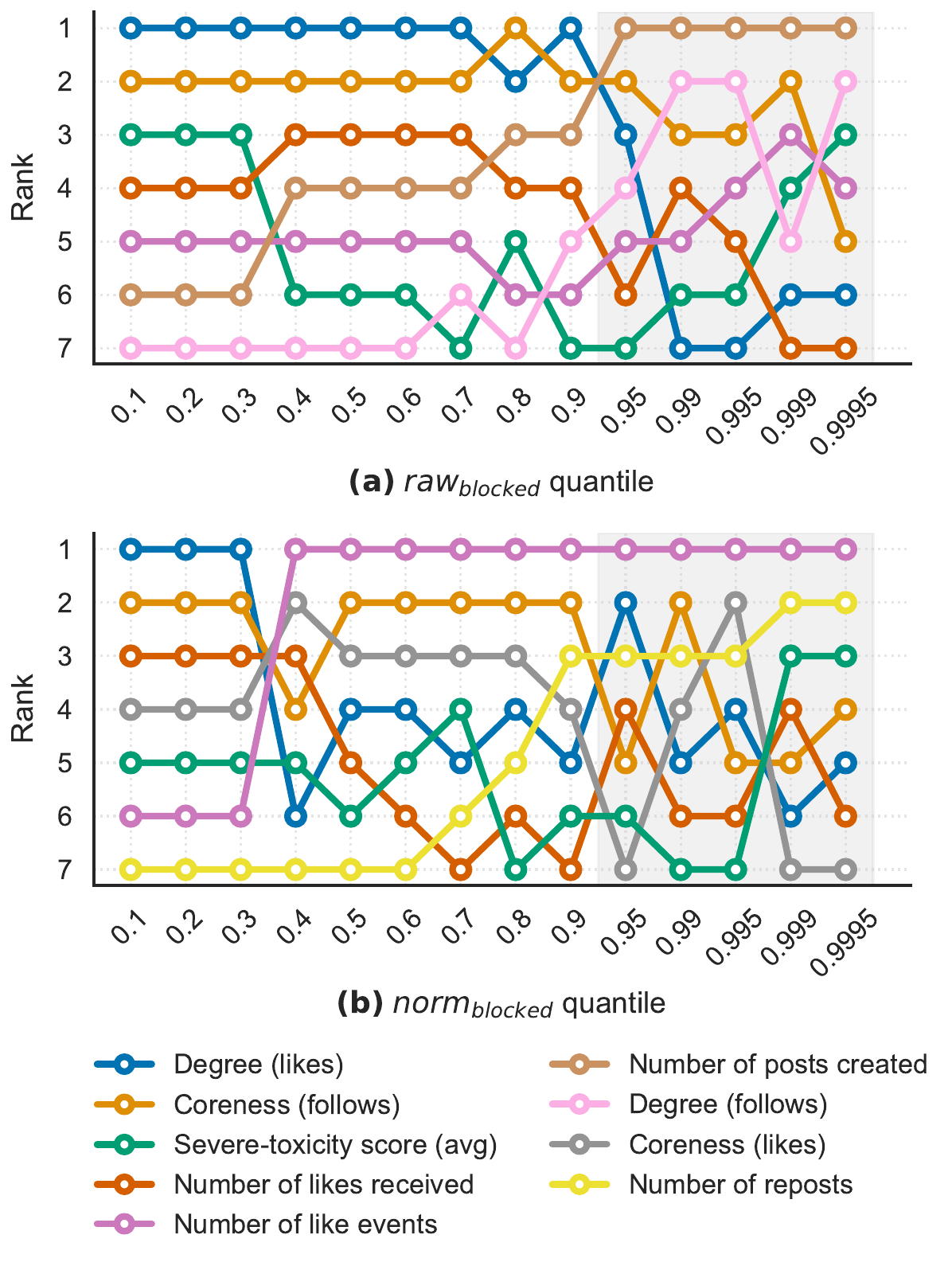} 
    \caption{Ranking of the union of the top 8 features, resulting in 9 unique features, by SHAP importance in the raw block count setting \textbf{(top)} and the normalized block count setting \textbf{(bottom)}, thresholding on progressive quantiles. }
    \label{fig:importance_bump}
\end{figure}

Since feature importance depends on how \textit{blocked} is defined, Figure~\ref{fig:importance_bump} reports the union of the top-ranked features across thresholds for both classification settings, yielding nine features within the top eight per setting, indicating strong overlap among the most informative signals. Some features that are important at lower thresholds lose relevance for heavily blocked users, such as the degree in the \textit{likes} graph and the number of \textit{likes} received. Conversely, other features gain importance at higher thresholds. In the raw block count setting (panel~\textbf{a}), this includes degree in the \textit{follows} graph, number of \textit{likes}, and number of \textit{posts}. In the normalized block count setting (panel~\textbf{b}), the most discriminative features are the number of \textit{likes}, number of \textit{reposts}, and the \textit{toxicity} score. These patterns suggest that frequently blocked users exhibit distinctive behavioral fingerprints, consistent with the classification results discussed in the previous section.

\begin{figure}[!t]
    \centering
    \includegraphics[width=\linewidth]{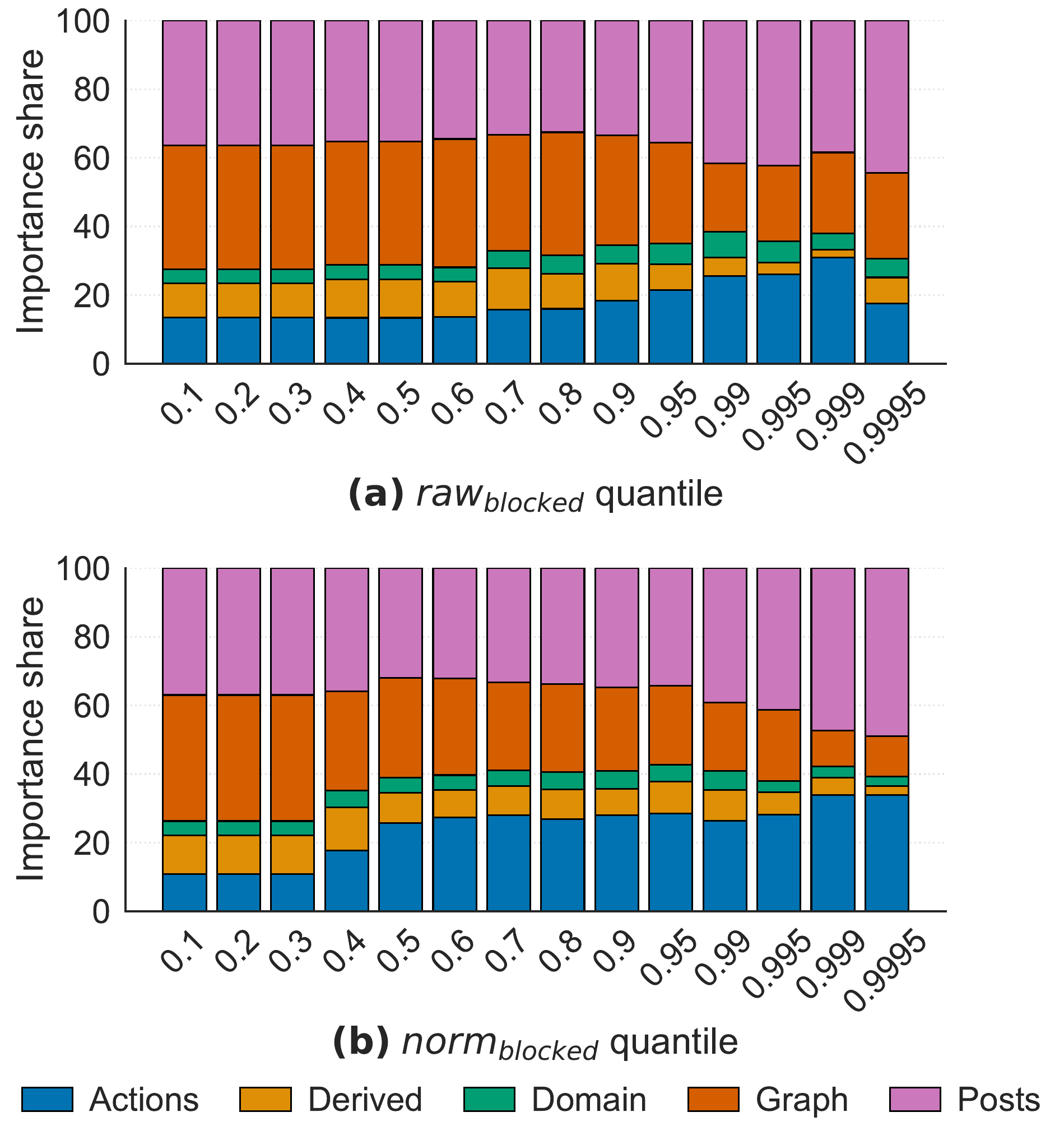} 
    \caption{SHAP importance values aggregated by feature group across thresholds in the raw \textbf{(a)} and normalized block count setting \textbf{(b)}. Labels in the legend indicate groups of features as defined in the Methods.}
    \label{fig:shap_group}
\end{figure}

Figure~\ref{fig:shap_group} shows how feature importance at the group level varies with the threshold. Both settings exhibit similar patterns: Posts features are the most important groups on average (respectively $\approx\!37\%$ and $\approx\!28\%$). However, at higher thresholds, Graph features appear to be less discriminating. 
At the same time, as blocking behavior becomes more pronounced, Actions gain relative importance and Domain and Derived features become less relevant. This indicates, overall, that activity patterns and content are strong indicators of frequent block targets, especially for extreme users. 


\begin{figure}[!t]
    \centering
    \includegraphics[width=\linewidth]{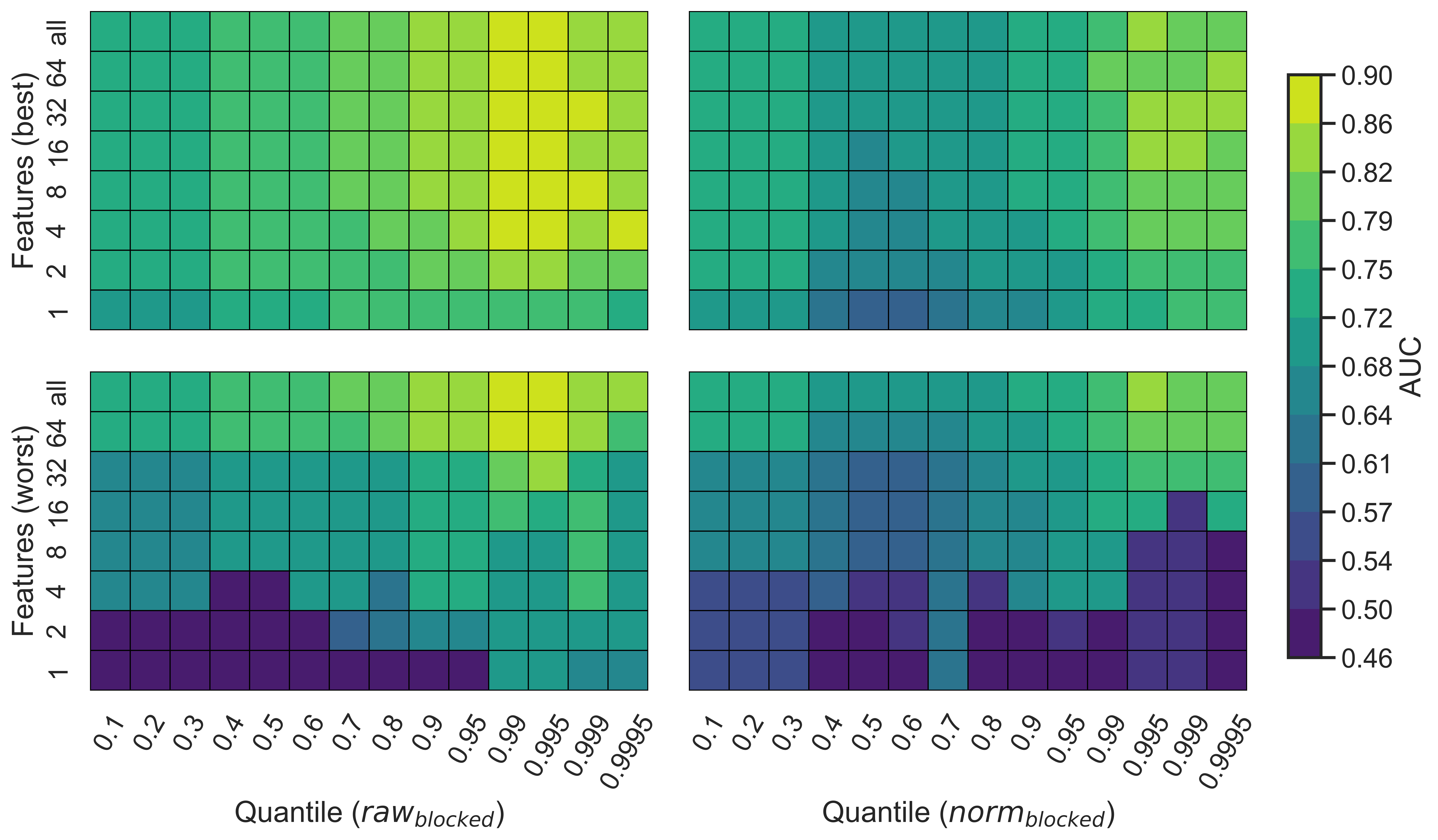} 
    \caption{Classification performance with increasing numbers of features. The \textbf{top} row shows classifiers trained on the best $n$ features (ranked by importance), and the \textbf{bottom} row on the worst $n$ features; left and right columns correspond to the raw and normalized block count settings, respectively.}
    \label{fig:heatmap_combo}
\end{figure}

The classification results of the feature ablation experiment are reported in Figure~\ref{fig:heatmap_combo}, which contains four panels: the top row shows classifiers trained on the \textit{best} $n$ features (ranked by SHAP importance), and the bottom row shows classifiers trained on the \textit{worst} $n$ features; the left and right columns correspond to the raw and normalized block count settings, respectively. Ten classifiers are trained on independent random samples, and the average ROC AUC is reported. In the top row, results show that a limited set of features can achieve reasonable performance. At higher thresholds, as few as four features are sufficient to achieve near-maximum performance, with additional features contributing only marginally. In the bottom row, the opposite pattern can be observed: using only a few of the least informative features yields performance close to random classification. For both low and high thresholds, performance comparable to that of the top two features requires including at least 64 of the worst features. These results highlight the value of feature selection: a small subset of informative features is sufficient for robust predictions, while adding less informative features provides little benefit. At the same time, the degree of informativeness depends on the definition of \textit{blocked} users.



\subsection{Regression analysis}
\begin{figure}[!t]
    \centering
    \includegraphics[width=\linewidth]{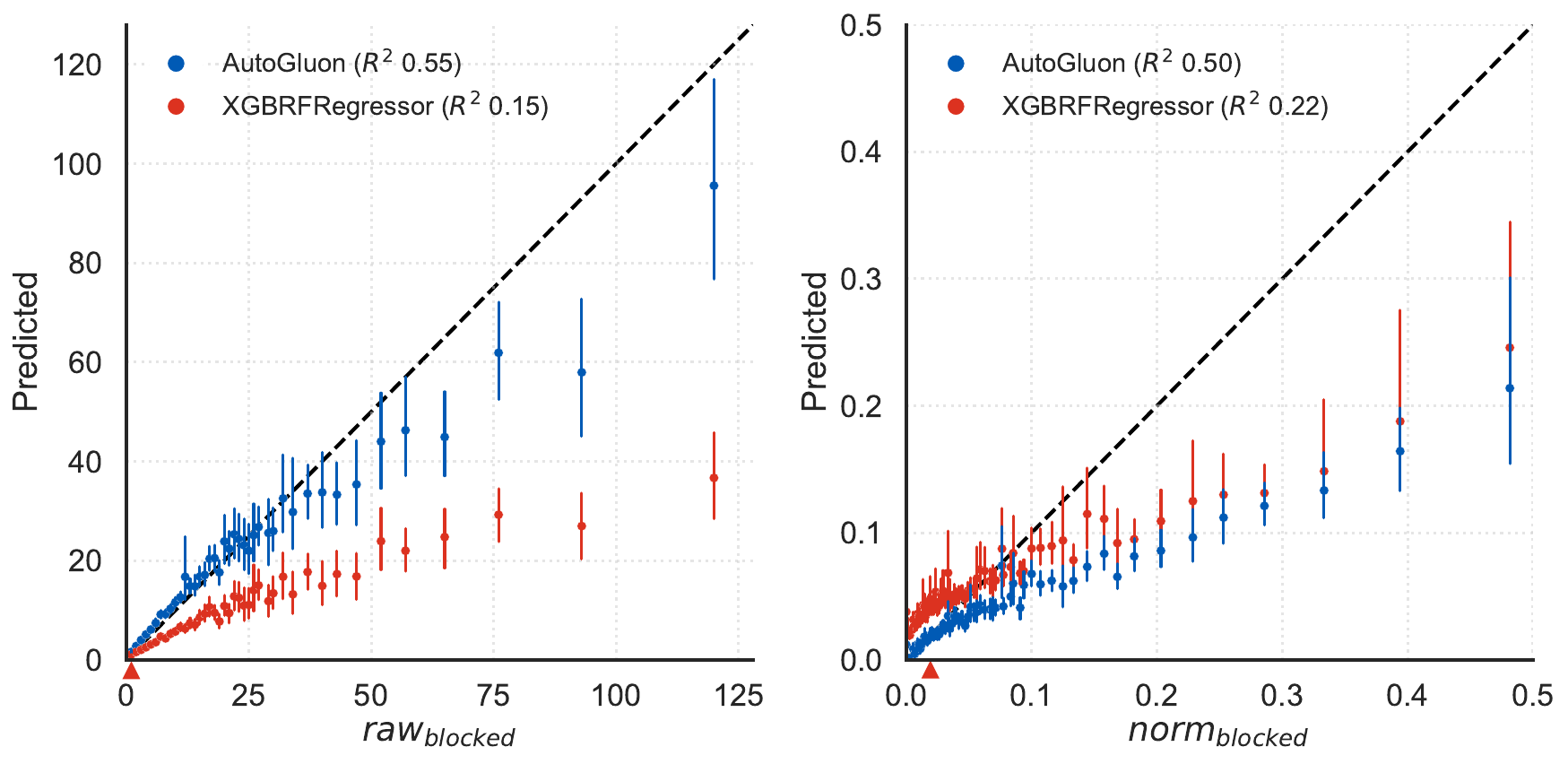}
    \caption{\textbf{(left)} Regression of raw block count per user, and \textbf{(right)} normalized block count, comparing XGBRFRegressor (\textcolor{xgbred}{red}) and AutoGluon (\textcolor{autogluonblue}{blue}). Markers correspond to percentile bins of the true values, the red triangle denotes the median.}
    \label{fig:regression}
\end{figure}

Given the strong classification performance, particularly for frequently blocked users, we next assess whether regression models can estimate the actual number of blocks received.
Figure~\ref{fig:regression} reports results for the two targets --- $raw_{blocked}$ (left) and $norm_{blocked}$ (right) block counts --- using Random Forest Regression (\textcolor{xgbred}{red}) and AutoGluon (\textcolor{autogluonblue}{blue}). 

Across both settings, AutoGluon consistently generalizes better than the baseline Random Forest model, though at a higher computational cost (about one hour of training versus under one minute for the baseline). 
Prediction accuracy is higher for users with fewer blocks, while both models underperform for the most heavily blocked users, a comparatively rare group, where estimates of $raw_{blocked}$ and $norm_{blocked}$ deviate more substantially from the true values. 
Nevertheless, the median values of both variables are very low, meaning that most users fall within the range where predictions are more reliable. To illustrate this practically, approximately 84\% of users with zero $raw_{blocked}$ blocks are correctly predicted to receive fewer than one block.  

\section{Discussion}
\label{sec:conclusions}
\subsubsection{Contributions}
This work presents the first large-scale longitudinal study of blocking behavior on Bluesky, based on more than 3 million block actions carried out by nearly 2 million users over a three-month period. 
We develop an interpretable, ML-based framework for retrospective identification of blocking outcomes, providing a foundation for future early-warning models and for causal investigations into the effects of blocking on user activity.

Our descriptive analysis shows that blocks, a less frequent interaction compared to likes, follows, and posts, are largely employed by Bluesky users and follow a heavy-tailed distribution. 
A small group of highly active or potentially misbehaving accounts stands out with extreme block-receiving activity. 
We also find that blocked users tend to be more active and moderately more toxic, though blocking does not appear to be systematically linked to sharing low-credibility content.  

Building on these insights, we investigated the extent to which blocked and non-blocked users can be distinguished. 
We framed this as a binary classification task using two target definitions: raw block counts and activity-normalized block counts. 
We then used quantiles of their respective distributions as classification thresholds. 
A Random Forest classifier consistently achieved strong performance, with more extreme users being the easiest to identify, particularly in the normalized setting, suggesting that being blocked is associated with distinctive behavioral fingerprints.
However, because interpretability is closely linked to model performance, insights for users in the lower block quantiles, where predictive accuracy is more limited, should be treated with caution. 
At the same time, the consistency of feature-importance patterns across thresholds provides additional evidence of robustness.

We further examined feature importance at group and individual levels using SHAP values, which provide locally consistent, model-agnostic explanations of complex ensemble predictions and are widely used in social computing research to interpret behavioral and content-based models~\cite{shevtsov2022identification,mathew2019thou,yang2020scalable}.
Results highlight how the relative importance of highly informative features shifts depending on the definition of \textit{blocked} and the selected threshold. 
Moreover, we show that graph-based features are more predictive at lower thresholds, while posting activity and content become more important for heavily blocked users. 
We also discuss how high classification performance can be achieved using fewer features. 

Finally, we modeled the number of blocks received as a regression task. 
AutoGluon outperformed Random Forest Regression in both raw and normalized settings, though at a greater computational cost. Both models perform reliably for the majority of users with few blocks but struggle to capture extreme cases. However, given that the median block counts are very low, this limitation is not a major concern in practice.  

Our work draws on research in moderation, decentralized governance, and self-moderation dynamics.
While prior studies~\cite{vogels2021state,seering2020reconsidering,kaiser2022partisan,martel2024blocking,baysha2020dividing,ashkinaze2024dynamics,hunt2018no,fox2015dark,zhang2020privacy} have examined user-driven moderation behaviors such as muting or unfollowing on platforms like Twitter, our work is the first to analyze behavioral patterns in relation to blocking on Bluesky at scale, enabled by the platform’s unique data openness.
By framing blocking as both a classification and a regression task, we show that behavioral traces can retrospectively predict blocking with meaningful accuracy, providing a methodological bridge toward future early-warning systems and causal analyses of how blocking shapes user behavior. 
In doing so, our study demonstrates how decentralized transparency can advance evidence-driven moderation research and practice.

\subsubsection{Implications}
There are several implications of our findings.
First, our study highlights that self-moderation in the form of user blocking is a widely used mechanism on Bluesky, suggesting that users actively shape their online environments by filtering interactions. 
We demonstrate that users who receive a high number of blocks exhibit distinctive behavioral traits that set them apart from the general user population, even when accounting for higher activity. 

Second, these distinctive traits can be effectively encoded and leveraged by machine learning models, demonstrating not only that blocking outcomes are retrospectively predictable but also that such predictability provides a foundation for future early-warning or flagging systems that could help moderation teams surface potentially problematic users before issues escalate. 
Notably, comparable performance can be achieved with parsimonious models that rely on only a handful of features, opening avenues for lightweight, privacy-preserving implementations and for future work aimed at understanding the causal pathways linking user behavior and blocking outcomes.

An important implication of our study lies in the value of data transparency. 
Our analysis was made possible by the open nature of Bluesky’s data ecosystem, which provides access to behavioral and moderation-related data. 
Other major social platforms also employ blocking and similar self-moderation mechanisms, yet such data are typically inaccessible to researchers, even on other decentralized platforms like Mastodon.
Broader data sharing would significantly advance the scientific understanding of online social dynamics and enable collaborative, evidence-based approaches to moderation. 
Platforms could benefit from shared insights into behavioral markers of problematic interactions, while researchers and policymakers could better assess the effectiveness and consequences of moderation strategies. 
In this sense, Bluesky serves as a compelling case study for how openness can catalyze innovation in trust and safety research.

\subsubsection{Limitations}

The analysis has been conducted over a three-month period, which may not capture long-term trends or account for seasonal variations in user behavior. A limited time frame could miss important shifts in blocking activity that might emerge over a longer observation period. Moreover, we only analyze users' activity from a static perspective, i.e., by aggregating their behaviour over the entire period of analysis.

Our study focused on a large but simple set of behavioral features to analyze blocking activity, potentially overlooking other influential factors. Incorporating more advanced features, such as text embeddings from social media posts or graph embeddings representing user interaction networks, could enhance the predictive power of the models. Alternative toxicity assessment strategies could also be explored, differentiating multiple information sources (e.g., replies and reposts) and considering additional languages.

While this study examines the patterns and characteristics of users who are blocked, it does not explore the underlying motivations behind these actions. Understanding why users block others or get blocked --- whether due to harassment, misinformation, or other factors --- could provide more context and actionability for the findings. 

Lastly, due in part to technical constraints and in part to the current demographics of the social network, this study is limited to users who predominantly interact in the English language, which may not fully represent global user behavior.
Language, cultural norms, and social dynamics can significantly influence how users interact and block others.
Moreover, the Bluesky user base itself is not necessarily representative of other platforms or countries, limiting the generalizability of our findings beyond this specific context.

\subsubsection{Future Work}
A key area for future research is related to the underlying motivations behind user blocking behavior. 
While our study focused on estimating the number of blocks received, it did not investigate the reasons users engage in blocking actions. Understanding the psychological, social, or contextual factors that drive users to block others --- such as harassment, misinformation, or disagreements --- could provide insights into the root causes of blocking behavior.
A promising direction would be to incorporate data from Bluesky’s Labelers, community- or service-run classifiers that assign safety, content, or policy-relevant labels to posts and accounts, including the one operated by the platform itself. 
These labels could provide valuable ground truth signals, complementary to toxicity and media-bias annotations, enriching our analysis.
A qualitative analysis of the most frequently blocked users could reveal more about who these users are and the specific use cases --- such as harassment, misinformation, or conflict patterns --- that quantitative features alone cannot capture, informing more targeted moderation strategies.
This would help develop more accurate models of user behavior and enable platform developers to create more sophisticated systems for managing toxic or disruptive interactions.

Our models demonstrate that it is possible to estimate users’ overall propensity to be blocked based on their behavioral features. However, we do not attempt to predict the exact number of blocks a user may receive in the future based on past activity; this remains an avenue for future work. Moreover, in the current formulation, features are not normalized by account age, meaning that users who joined during the observation window may have shorter behavioral histories; incorporating appropriate countermeasures could further refine future modeling efforts.

Another direction for future research involves applying link prediction techniques to understand the network dynamics associated with blocking behavior. Link prediction aims to forecast the likelihood of future connections among specific users, based on their past interactions and respective positions within the network structure. By extending the proposed approach, we could predict which users are more likely to block or be blocked by others in the future. This would provide a deeper understanding of social networks and interactions on platforms, allowing for proactive identification of potential conflicts before they escalate into blocking actions, ultimately improving user experience.

Finally, future research would benefit from developing a causal framework to understand the effects of blocking on user behavior. While our study identifies patterns associated with users who block others, it does not establish a causal relationship between blocking actions and the subsequent impact on user interactions or overall platform engagement. Investigating how blocking influences users’ behavior --- changes in their activity, sentiment, or social connections --- might shed light on the broader consequences of blocking. This would enable platform designers to implement more effective tools and interventions to mitigate the negative effects of blocking while fostering healthier user interactions.

\bibliography{bib,aaai2026}

\newpage

\onecolumn
\clearpage 
\section*{Appendix}
\addcontentsline{toc}{section}{Appendix} 

\subsection*{Feature Descriptions}
Table~\ref{table:features} provides a complete list of the 86 behavioral features used in this work, computed at the user level.

\setlength{\tabcolsep}{4pt}

\begin{longtable}{>{\centering\arraybackslash}p{1.5cm} >{\centering\arraybackslash}p{0.8cm} l p{8cm}}
\caption{Description of the behavioral features, grouped by type.} \\
\toprule
\textbf{Group} & \textbf{\#} & \textbf{Feature} & \textbf{Description} \\
\midrule
\endfirsthead

\multicolumn{4}{c}{{\bfseries \tablename\ \thetable{} -- continued}} \\
\toprule
\textbf{Group} & \textbf{\#} & \textbf{Feature} & \textbf{Description} \\
\midrule
\endhead

\midrule \multicolumn{4}{r}{{Continued on next page}} \\
\endfoot

\bottomrule
\endlastfoot

\newcounter{rownum}
\setcounter{rownum}{1}

\multirow{16}{*}{\rotatebox{90}{Action / Derived}}
& \arabic{rownum} & create.app.bsky.feed.like        & Number of like events created \\ \stepcounter{rownum}
& \arabic{rownum} & create.app.bsky.feed.post        & Number of post events created \\ \stepcounter{rownum}
& \arabic{rownum} & create.app.bsky.feed.repost      & Number of repost events created \\ \stepcounter{rownum}
& \arabic{rownum} & create.app.bsky.graph.follow     & Number of follow events created \\ \stepcounter{rownum}
& \arabic{rownum} & delete.app.bsky.feed.like        & Number of like events deleted \\ \stepcounter{rownum}
& \arabic{rownum} & delete.app.bsky.feed.post        & Number of post events deleted \\ \stepcounter{rownum}
& \arabic{rownum} & delete.app.bsky.feed.repost      & Number of repost events deleted \\ \stepcounter{rownum}
& \arabic{rownum} & delete.app.bsky.graph.follow     & Number of follow events deleted \\ \stepcounter{rownum}
& \arabic{rownum} & derived.followed                 & Number of users following a user\\ \stepcounter{rownum}
& \arabic{rownum} & derived.follower                 & Number of users followed by a user \\ \stepcounter{rownum}
& \arabic{rownum} & derived.liked                    & Number of likes received \\ \stepcounter{rownum}
& \arabic{rownum} & derived.liker                    & Number of likes assigned \\ \stepcounter{rownum}
& \arabic{rownum} & derived.replied\_parent           & Number of replies received \\ \stepcounter{rownum}
& \arabic{rownum} & derived.replier\_parent           & Number of times replied \\ \stepcounter{rownum}
& \arabic{rownum} & derived.reposted                  & Number of reposts received \\ \stepcounter{rownum}
& \arabic{rownum} & derived.reposter                  & Number of times reposted  \stepcounter{rownum} \\

\midrule

\multirow{30}{*}{\rotatebox{90}{Post-derived}}
& \arabic{rownum} & posts.count                      & Total number of posts (supported languages) \\ \stepcounter{rownum}
& \arabic{rownum} & posts.language\_entropy           & Entropy of post languages \\ \stepcounter{rownum}

& \arabic{rownum} & posts.num\_chars\_mean            & Mean characters per post \\ \stepcounter{rownum}
& \arabic{rownum} & posts.num\_chars\_std             & Std characters per post \\ \stepcounter{rownum}
& \arabic{rownum} & posts.num\_digits\_mean           & Mean digits per post \\ \stepcounter{rownum}
& \arabic{rownum} & posts.num\_digits\_std            & Std digits per post \\ \stepcounter{rownum}
& \arabic{rownum} & posts.num\_emoji\_mean            & Mean emojis per post \\ \stepcounter{rownum}
& \arabic{rownum} & posts.num\_emoji\_std             & Std emojis per post \\ \stepcounter{rownum}
& \arabic{rownum} & posts.num\_lowercase\_mean        & Mean lowercase characters \\ \stepcounter{rownum}
& \arabic{rownum} & posts.num\_lowercase\_std         & Std lowercase characters \\ \stepcounter{rownum}
& \arabic{rownum} & posts.num\_uppercase\_mean        & Mean uppercase characters \\ \stepcounter{rownum}
& \arabic{rownum} & posts.num\_uppercase\_std         & Std uppercase characters \\ \stepcounter{rownum}
& \arabic{rownum} & posts.num\_spaces\_mean           & Mean spaces per post \\ \stepcounter{rownum}
& \arabic{rownum} & posts.num\_spaces\_std            & Std spaces per post \\ \stepcounter{rownum}

& \arabic{rownum} & posts.identity\_attack\_mean      & Mean ``Identity attack'' score (Detoxify) \\ \stepcounter{rownum}
& \arabic{rownum} & posts.identity\_attack\_std       & Std ``Identity attack'' score (Detoxify)\\ \stepcounter{rownum}
& \arabic{rownum} & posts.insult\_mean                & Mean ``Insult'' score (Detoxify)\\ \stepcounter{rownum}
& \arabic{rownum} & posts.insult\_std                 & Std ``Insult'' score (Detoxify)\\ \stepcounter{rownum}
& \arabic{rownum} & posts.obscene\_mean               & Mean ``Obscene'' score (Detoxify)\\ \stepcounter{rownum}
& \arabic{rownum} & posts.obscene\_std                & Std ``Obscene'' score (Detoxify)\\ \stepcounter{rownum}
& \arabic{rownum} & posts.threat\_mean                & Mean ``Threat'' score (Detoxify)\\ \stepcounter{rownum}
& \arabic{rownum} & posts.threat\_std                 & Std ``Threat'' score (Detoxify)\\ \stepcounter{rownum}
& \arabic{rownum} & posts.sexual\_explicit\_mean      & Mean ``Sexual explicit'' score (Detoxify)\\ \stepcounter{rownum}
& \arabic{rownum} & posts.sexual\_explicit\_std       & Std ``Sexual explicit'' score (Detoxify)\\ \stepcounter{rownum}
& \arabic{rownum} & posts.severe\_toxicity\_mean      & Mean ``Severe toxicity'' score (Detoxify)\\ \stepcounter{rownum}
& \arabic{rownum} & posts.severe\_toxicity\_std       & Std ``Severe toxicity'' score (Detoxify)\\ \stepcounter{rownum}
& \arabic{rownum} & posts.toxicity\_mean              & Mean ``Toxicity score'' (Detoxify)\stepcounter{rownum} \\
& \arabic{rownum} & posts.toxicity\_std               & Std ``Toxicity score'' (Detoxify)\stepcounter{rownum} \\

\pagebreak

\multirow{25}{*}{\rotatebox{90}{Domain / URL-derived}}
& \arabic{rownum} & domain.url\_count                  & Average number of URLs per post \\ \stepcounter{rownum}
& \arabic{rownum} & domain.urls\_entropy               & Domain entropy \\ \stepcounter{rownum}
& \arabic{rownum} & domain.bias\_entropy              & Bias value entropy (MBFC)\\ \stepcounter{rownum}
& \arabic{rownum} & domain.bias\_center               & Fraction of ``center'' domains (MBFC)\stepcounter{rownum} \\ 
& \arabic{rownum} & domain.bias\_conspiracy           & Fraction of ``conspiracy'' domains (MBFC)\stepcounter{rownum} \\
& \arabic{rownum} & domain.bias\_extreme-left         & Fraction of ``extreme-left'' domains (MBFC)\\ \stepcounter{rownum}
& \arabic{rownum} & domain.bias\_extreme-right        & Fraction of ``extreme-right'' domains (MBFC)\\ \stepcounter{rownum}
& \arabic{rownum} & domain.bias\_left                 & Fraction of ``left'' domains (MBFC)\\ \stepcounter{rownum}
& \arabic{rownum} & domain.bias\_left-center          & Fraction of ``left-center'' domains (MBFC)\\ \stepcounter{rownum}
& \arabic{rownum} & domain.bias\_right                & Fraction of ``right'' domains (MBFC) \\ \stepcounter{rownum}
& \arabic{rownum} & domain.bias\_right-center         & Fraction of ``right-center'' domains (MBFC) \\ \stepcounter{rownum}
& \arabic{rownum} & domain.bias\_satire               & Fraction of ``satire'' domains (MBFC) \\ \stepcounter{rownum}
& \arabic{rownum} & domain.bias\_pro-science          & Fraction of ``pro-science'' domains (MBFC) \\ \stepcounter{rownum}
& \arabic{rownum} & domain.credibility\_high          & Fraction of ``high credibility'' domains (MBFC)\\ \stepcounter{rownum}
& \arabic{rownum} & domain.credibility\_medium        & Fraction of ``medium credibility'' domains (MBFC)\\ \stepcounter{rownum}
& \arabic{rownum} & domain.credibility\_low           & Fraction of ``low credibility'' domains (MBFC)\\ \stepcounter{rownum}
& \arabic{rownum} & domain.factual\_very\_high        & Fraction of ``very high factual'' domains (MBFC)\\ \stepcounter{rownum}
& \arabic{rownum} & domain.factual\_high              & Fraction of ``high factual'' domains (MBFC)\\ \stepcounter{rownum}
& \arabic{rownum} & domain.factual\_mostly            & Fraction of ``mostly factual'' domains (MBFC)\\ \stepcounter{rownum}
& \arabic{rownum} & domain.factual\_mixed             & Fraction of ``mixed factual'' domains (MBFC)\\ \stepcounter{rownum}
& \arabic{rownum} & domain.factual\_low               & Fraction of ``low factual'' domains (MBFC)\\ \stepcounter{rownum}
& \arabic{rownum} & domain.factual\_very\_low          & Fraction of ``very low factual'' domains (MBFC)\\ \stepcounter{rownum}
& \arabic{rownum} & domain.bias\_count                 & Count of domains with a bias value (MBFC) \\ \stepcounter{rownum}
& \arabic{rownum} & domain.credibility\_count          & Count of domains with a Credibility value (MBFC) \\ \stepcounter{rownum}
& \arabic{rownum} & domain.factual\_count              & Count of domains with a Factual value (MBFC) \\ \stepcounter{rownum}
& \arabic{rownum} & domain.pc1\_count                  & Domain quality score count \cite{Lin2023Sep}\\ \stepcounter{rownum}
& \arabic{rownum} & domain.pc1\_mean                   & Domain quality score mean \cite{Lin2023Sep}\stepcounter{rownum} \\

\midrule

\multirow{15}{*}{\rotatebox{90}{Graph-based}}
& \arabic{rownum} & graph.follows\_coreness            & Coreness in follows network \\ \stepcounter{rownum}
& \arabic{rownum} & graph.follows\_degree              & Degree in follows network \\ \stepcounter{rownum}
& \arabic{rownum} & graph.follows\_pagerank            & PageRank in follows network \\ \stepcounter{rownum}
& \arabic{rownum} & graph.likes\_coreness              & Coreness in likes network \\ \stepcounter{rownum}
& \arabic{rownum} & graph.likes\_degree                & Degree in likes network \\ \stepcounter{rownum}
& \arabic{rownum} & graph.likes\_pagerank              & PageRank in likes network \\ \stepcounter{rownum}
& \arabic{rownum} & graph.likes\_strength              & Weighted degree in likes network \\ \stepcounter{rownum}
& \arabic{rownum} & graph.replies\_parent\_coreness    & Coreness in replies network \\ \stepcounter{rownum}
& \arabic{rownum} & graph.replies\_parent\_degree      & Degree in replies network \\ \stepcounter{rownum}
& \arabic{rownum} & graph.replies\_parent\_pagerank    & PageRank in replies network \\ \stepcounter{rownum}
& \arabic{rownum} & graph.replies\_parent\_strength    & Weighted strength in replies network \\ \stepcounter{rownum}
& \arabic{rownum} & graph.reposts\_coreness             & Coreness in reposts network \\ \stepcounter{rownum}
& \arabic{rownum} & graph.reposts\_degree               & Degree in reposts network \\ \stepcounter{rownum}
& \arabic{rownum} & graph.reposts\_pagerank             & PageRank in reposts network \\ \stepcounter{rownum}
& \arabic{rownum} & graph.reposts\_strength             & Weighted strength in reposts network\stepcounter{rownum} 

\label{table:features}
\end{longtable}

\end{document}